\documentclass[aps, prd ]{revtex4}

\usepackage[small]{caption}
\usepackage{amsmath,amsfonts,amscd,amssymb,epsf, epsfig,mathrsfs,graphicx}

\newcommand{\vep}{\varepsilon}

\begin{document}

\title{ First analytic correction to the proximity force approximation in the Casimir effect between two parallel cylinders}

 \author{L. P. Teo}\email{LeePeng.Teo@nottingham.edu.my}
\address{Department of Applied Mathematics, Faculty of Engineering, University of Nottingham Malaysia Campus, Jalan Broga, 43500, Semenyih, Selangor Darul Ehsan, Malysia.}
 \begin{abstract}
We consider the small separation asymptotic expansions of the Casimir interaction energy and the Casimir interaction force between two parallel cylinders. The leading order terms and the next-to-leading order terms are computed analytically. Four combinations of boundary conditions are considered, which are Dirichlet-Dirichlet (DD), Neumann-Neumann (NN), Dirichlet-Neumann (DN) and Neumann-Dirichlet (ND). For the case where one cylinder is inside another cylinder, the computations are shown in detail. In this case, we restrict our attention to the situation where the cylinders are strictly eccentric and  the distance between the cylinders $d$ is much smaller than the distance between the centers of the cylinders. The computations for the case where the two cylinders are exterior to each other can be done in the same way and we only present the results, which turn up to be similar to the results for the case where one cylinder is inside another except for some changes of   signs.  In all the scenarios we consider, the leading order terms are of order $d^{-7/2}$ and they agree completely with the proximity force approximations. The results for the next-to-leading order terms are new. In the limiting case where the radius of the larger cylinder approaches infinity, the well-known results for  the cylinder-plate configuration with DD or NN boundary conditions are recovered.
 \end{abstract}

\maketitle
\section{Introduction}
Recently, there has been an increasing interest in the Casimir effect from both the theoretical and the experimental sides \cite{18}. Before the turn of this century, the theoretical studies on the exact Casimir effect were mostly restricted to simple geometries such as parallel plates,   spherical shells and   cylindrical shells. However,   these geometric configurations pose a certain degree of difficulty for the experimental verification of
the Casimir effect, such as the difficulty in achieving parallelism. As a result, experimentalists favor other configurations, especially the sphere-plane configuration. However, in the last century, one has to rely on the proximity force approximation to estimate the Casimir force between such configurations, which hampers the determination of the experimental accuracies. To circumvent the problem, intensive activities have been carried out to research for theoretical methods to determine the  Casimir interactions between two or several objects beyond the accuracies afforded by the proximity force approximations. In the last decade, a number of methods have been developed, which include the  semi-classical approximation \cite{19,20}, the optical path method \cite{21,22,23}, the worldline approach \cite{24,25,26,27}, the functional determinant or the multiple scattering method \cite{10,28,29,30,31,32,33,34,51,4}, and the exact mode summation method \cite{2,3}. Using the multiple scattering method, one can in principle write down a functional for the Casimir interaction energy between two or several objects. Nevertheless, it should be mentioned that the mode summation approach can also lead to the same result, although this latter method has only been applied to the configuration of two eccentric cylinders.

In principle, using the various methods mentioned above, one can compute the magnitude of the Casimir force numerically. However, the accuracy is always subjected to the computing capacity of the computer, especially when the separation between the objects is small, which is of more interest for comparison to experiments. On the other hand, to determine the dependence of the Casimir force on various parameters of a configuration, one usually has to assume a priori the form of the dependence  on the parameters and determine the coefficients that fit best for that particular form. These coefficients are subjected to numerical errors and it is not easy to justify the accuracy of the results. Therefore there is a call for analytically computing the asymptotic expansion of the Casimir force when the separation between the objects is small. For the cylinder-plate and the sphere-plate configurations, the first corrections to the proximity force approximations have been computed analytically in \cite{10,12,13,35}. Although the  configurations of two cylinders and two spheres are among the popular ones whose exact Casimir interaction energies have been derived using the multiple scattering approach or the mode summation method \cite{28,30,31,32,33,34,51,4,2,3,8,36,37},   analytical studies on the corrections to the proximity force approximations for these configurations are still  lacking. The purpose of this article is to address this problem for the configuration of two cylinders. The results would be compared to the results for the cylinder-plate configuration which is a special case where the radius of one of the cylinders approaches infinity.

In \cite{8,1},  some experimental setups have been proposed to measure the Casimir interaction force between two eccentric cylinders.  It was argued that the cylindrical configuration has some of the experimental advantages
of both the the parallel planes and the  sphere-plane configurations. It can lead to favorable conditions
to   search for extra-gravitational forces in the micrometer range and for the observation
of finite-temperature corrections. These latter subjects have  been explored for the cylinder-plane configuration in \cite{38,39,40,41}. In view of this, the cylindrical configuration is becoming increasingly important in the study of the Casimir effect. It is timely to do an analytical study on the strength of the Casimir force when the separation between the cylinders is small.

In this article, we  consider   the case  where one cylinder is inside another, and the case where two cylinders are outside each other.  The functional determinant representations  of the exact Casimir interaction energies have been derived in several places, such as \cite{2,3,6,17,4,5} for the case of one cylinder inside another, and \cite{17,4,5,18} for the case of two cylinders exterior to each other. Using this formulas, we compute analytically the first correction to the proximity force approximations. In the case of one cylinder inside another, the roles of the two cylinders are not symmetrical. We explain the computations  in detail for this case. The computations for the case of two cylinders outside each other can be done in the same way and we only present the results.

Throughout the paper, we use the units with $\hbar=c=1$.
\section{Proximity force approximation of the Casimir interaction between two  parallel eccentric cylinders}\label{s2}
\begin{figure}[h]\centering
\epsfxsize=0.5\linewidth \epsffile{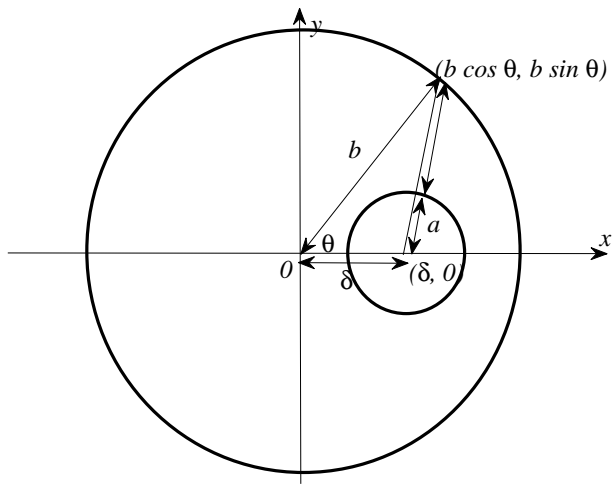}  \caption{\label{f1} The cross section of two eccentric cylinders.}  \end{figure}
As shown in Fig. \ref{f1}, we consider two parallel cylinders of length $L$ and radii $a$ and $b$ respectively. The cylinder of radius $a$ lies inside the cylinder of radius $b$. Denote by $\delta$ the separation between the centers of the cylinders, and $d$ the distance between the cylinders. Clearly, we have $\delta=b-a-d$.   In this article, we assume exclusively $\delta>0$, i.e., the cylinders are not concentric. The concentric case has been considered in \cite{7,8,9}. For the boundary conditions on the cylinders, we impose   Dirichlet or Neumann boundary conditions. In the following, we will denote the boundary conditions by XY, where $\text{X}=\text{D}$ (Dirichlet) or N (Neumann) is the boundary condition on the cylinder of radius $a$, and Y $=$ D or N for the cylinder of radius $b$.

In this section, we use the proximity force approximation (PFA) to obtain the leading term of the Casimir interaction force between the two cylinders when $d\ll \delta\ll b-a$. Notice that the radii $a$ and $b$ are fixed parameters, whereas the distance between the cylinders $d$ is a variable.

First, recall that the Casimir interaction force per unit area on two parallel plates separated by a distance $H$ is given by
\begin{equation}\label{eq7_28_2}\mathcal{F}^{\parallel}_{\text{Cas}}(H)=-\frac{\pi^2}{480H^4},\end{equation} if the two plates are both imposed with Dirichlet boundary conditions or Neumann boundary conditions. If one plate is imposed with Dirichlet boundary condition and one plate is imposed with Neumann boundary condition,  the Casimir interaction force density
is $-7/8$ times of \eqref{eq7_28_2}, i.e.,
\begin{equation}\mathcal{F}^{\parallel}_{\text{Cas}}(d)=\frac{7}{8}\frac{\pi^2}{480H^4}=\frac{7\pi^2}{3840 H^4}.\end{equation}
Using PFA, we have to integrate the Casimir energy density $\mathcal{F}^{\parallel}_{\text{Cas}}(H)$ over the area of one of the cylinders, with $H$ being the distance from a point of the integrated cylinder to the other cylinder. Here we choose to integrate over the cylinder of radius $b$. The integration over the length of the cylinder is trivial. Using polar coordinates, we find that the shortest distance from the point with parameter $\theta$ (see Fig. 1) to the cylinder of radius $a$ is
$$\sqrt{(b\cos\theta-\delta)^2+(b\sin\theta)^2}-a=\sqrt{b^2+\delta^2-2b\delta\cos\theta}-a.$$
Then for the DD or the NN case, the PFA for the Casimir interaction force between the cylinders is
\begin{equation*}\begin{split}
F_{\text{Cas}}^{\text{PFA}}= -\frac{\pi^2 b L}{240}\int_0^{\pi}\frac{1}{\left(\sqrt{b^2+\delta^2-2b\delta\cos\theta}-a\right)^4}d\theta.
\end{split}
\end{equation*} Making a change of variables
\begin{equation*}u=\frac{\sqrt{b^2+\delta^2-2b\delta\cos\theta}-a}{d},\end{equation*} we find that
\begin{equation*} \begin{split}\cos\theta=&\frac{b^2+\delta^2-(du+a)^2}{2b\delta},\\
\sin\theta=&\frac{\sqrt{4b^2\delta^2-[b^2+\delta^2-(du+a)^2]^2}}{2b\delta}=\frac{\sqrt{d(u-1)(2a+du+d)\left[(2b-a-d)^2-(du+a)^2\right]}}{2b\delta},\\
du=& \frac{b\delta\sin\theta}{d(du+a)}d\theta,
\end{split}\end{equation*}
 and thus
\begin{equation*}\begin{split}
F_{\text{Cas}}^{\text{PFA}}= -\frac{\pi^2 b  L}{240}\int_1^{\frac{2(b-a)-d}{d}}\frac{d(du+a)}{d^4u^4} \frac{2}{
 \sqrt{d(u-1)(2a+du+d)\left[(2b-a-d)^2-(du+a)^2\right]}}du.
\end{split}
\end{equation*}In the limit $d\rightarrow 0$, we find that the leading order term of the PFA is
\begin{equation}\label{eq7_28_1}\begin{split}
F_{\text{Cas}}^{\text{PFA}}\sim &-\frac{\pi^2 \sqrt{ab} L}{240\sqrt{2  (b-a)} d^{\frac{7}{2}}}\int_1^{\infty}\frac{1}{ u^4
 \sqrt{ u-1  }}du\\
=&-\frac{ \pi^3 \sqrt{ab} L}{768\sqrt{2  (b-a)} d^{\frac{7}{2}}}.
 \end{split}
\end{equation}For two perfectly conducting eccentric cylinders, the proximity force approximation to the Casimir interaction force is twice that of \eqref{eq7_28_1}, corresponding to the sum of the TE (which is equal to  NN) and the TM (which is equal to  DD) contributions. This has been obtained in \cite{1}.

For the DN or the ND case, one just has to multiply \eqref{eq7_28_1} by $-7/8$ which gives
\begin{equation}\begin{split}
F_{\text{Cas}}^{\text{PFA}}\sim &  \frac{7\pi^3 \sqrt{ab} L}{6144\sqrt{2  (b-a)} d^{\frac{7}{2}}}.
 \end{split}
\end{equation}

\section{The formula for the exact Casimir interaction energy between the parallel eccentric cylinders}\label{s3}
The exact Casimir interaction energy between two eccentric cylinders has been derived in \cite{2,3,6} using mode summation approach and in \cite{17,4,5} using scattering or functional determinant method.
  The Casimir interaction energy can be represented as
\begin{equation}\label{eq7_29_1}
E_{\text{Cas}} = \frac{L}{4\pi}\int_0^{\infty}\xi \text{Tr}\ln \left(\mathbf{1}-\mathbf{M}(\xi)\right)d\xi,
\end{equation}where $\mathbf{M}$ is an $\infty\times \infty$ matrix with elements $M_{mn}$, $-\infty<m,n<\infty$,
\begin{equation}\label{eq7_29_2}\begin{split}
M_{mn}^{\text{DD}}(\xi)=&\frac{I_n(  a\xi)}{K_{m}(  a \xi)}\sum_{p=-\infty}^{\infty}\frac{K_p(  b \xi)}{I_p(  b\xi)}I_{p-m}( \delta \xi)I_{p-n}(\delta\xi)\\
M_{mn}^{\text{NN}}(\xi)=&\frac{I_n'(  a\xi)}{K_{m}'(  a \xi)}\sum_{p=-\infty}^{\infty}\frac{K_p'(  b \xi)}{I_p'(  b\xi)}I_{p-m}( \delta \xi)I_{p-n}(\delta\xi),\\
M_{mn}^{\text{DN}}(\xi)=&\frac{I_n(  a\xi)}{K_{m}(  a \xi)}\sum_{p=-\infty}^{\infty}\frac{K_p'(  b \xi)}{I_p'(  b\xi)}I_{p-m}( \delta \xi)I_{p-n}(\delta\xi),\\
M_{mn}^{\text{ND}}(\xi)=&\frac{I_n'(  a\xi)}{K_{m}'(  a \xi)}\sum_{p=-\infty}^{\infty}\frac{K_p(  b \xi)}{I_p(  b\xi)}I_{p-m}( \delta \xi)I_{p-n}(\delta\xi).
\end{split}\end{equation}In these formulas, $I_{\nu}(z)$ and $K_{\nu}(z)$ are the modified Bessel functions of first and second kinds respectively. Recall that the dependence   on the distance $d$ between the cylinders is encoded in the variable $\delta=b-a-d$. From \eqref{eq7_29_2}, we notice that the terms dependent on $d$, $I_{p-m}( \delta \xi)I_{p-n}(\delta\xi)$, is identical for the four combinations of boundary conditions. The factors $I_{p-m}( \delta \xi)$ and $I_{p-n}(\delta\xi)$ actually come from translation formulas.

Expanding the logarithm and taking the trace in the formula \eqref{eq7_29_1}, we find that the Casimir interaction energy can be evaluated as
\begin{equation}\label{eq7_29_3}
E_{\text{Cas}} =- \frac{L}{4\pi}\sum_{s=0}^{\infty}\frac{1}{s+1}\int_0^{\infty}\xi \sum_{j_0=-\infty}^{\infty}\ldots\sum_{j_s=-\infty}^{\infty} M_{j_0j_1}(\xi)\ldots M_{j_sj_0}(\xi)d\xi.
\end{equation}Alternatively, one can also use the identity $\text{Tr}\ln =\ln \det$ to write the Casimir interaction energy \eqref{eq7_29_1}
as \begin{equation}\label{eq7_29_4}
E_{\text{Cas}} = \frac{L}{4\pi}\int_0^{\infty}\xi  \ln \det\left(\mathbf{1}-\mathbf{M}(\xi)\right)d\xi.
\end{equation}In principle, the magnitude of the Casimir interaction energy can be evaluated with the help of a computer. When the separation between the cylinders $d$ is much larger than the radius $a$ of the smaller cylinder, one can in fact determine analytically the dominating term of the Casimir interaction energy from \eqref{eq7_29_3}, since in this case, the dominating term is the term with $s=0$ and $j_0=0$. In the opposite limit where $d\ll a,b$, which is of more experimental interest, even the numerical computation of the magnitude of the Casimir interaction energy or the Casimir interaction force poses a great challenge since one has to take the matrix $\mathbf{M}$ with larger size in \eqref{eq7_29_4} for convergence. Some numerical results have been discussed in \cite{14} for the special cases of quasiconcentric cylinders (where $\delta\ll a$), concentric cylinders (where $\delta=0$) and the limiting case of a cylinder in front of a plane (where $b\rightarrow \infty$). In the following, we will  use analytical approach to compute the leading order term and the first order correction term of the Casimir interaction energy and the Casimir interaction force when $d\ll a,b$. In addition, we also assume that $d\ll b-a$. However, we do not make any assumption about the relative sizes of $\delta$ and $a$. As mentioned above, we assume exclusively that $\delta\neq 0$. In fact, $d\ll b-a$ implies $d\ll \delta$ and $\delta\sim b-a$.
The method we use is similar to that used in \cite{10,11} for a cylinder in front of a plane and in \cite{12,13,15} for a sphere in front of a plane. Compare to the case of a cylinder in front of a plane discussed in \cite{10,11}, the main complication here is the summation over $p$ that appears in the matrix elements $M_{mn}$ \eqref{eq7_29_2}.

\section{Asymptotic behavior of the Casimir interaction energy and the Casimir interaction force of the eccentric cylinders at small separation}\label{s4}
Define the dimensionless parameter
\begin{equation*}
\alpha=\frac{a}{b-a}
\end{equation*}and the dimensionless variable
\begin{equation*}
\vep=\frac{d}{b-a}.
\end{equation*}
Let $$\beta=\alpha+1=\frac{b}{b-a}.$$We are interested in the asymptotic behaviors of the Casimir interaction energy and the Casimir interaction force when $\vep\ll 1$.
Making a change of variables $\xi=\omega/(b-a)$ in \eqref{eq7_29_3}, we find that the Casimir interaction energy can be rewritten as
\begin{equation}\label{eq7_29_5}
E_{\text{Cas}} =- \frac{L}{4\pi (b-a)^2}\sum_{s=0}^{\infty}\frac{1}{s+1}\int_0^{\infty}\omega \sum_{j_0=-\infty}^{\infty}\ldots\sum_{j_s=-\infty}^{\infty} \sum_{p_0=-\infty}^{\infty}\ldots\sum_{p_s=-\infty}^{\infty} A_{j_0j_1;p_0}(\omega)\ldots A_{j_sj_0;p_s}(\omega)d\omega,
\end{equation}where $A_{mn;p}(\omega)=B_{mn;p}(\omega)T_{mn;p}(\omega)$,
\begin{equation}\label{eq7_29_6}\begin{split}
B_{mn;p}^{\text{DD}}(\omega)=&\frac{I_n(  \alpha\omega)}{K_{m}(  \alpha\omega)} \frac{K_p(  \beta\omega)}{I_p(  \beta\omega)},\hspace{1cm}
B_{mn;p}^{\text{NN}}(\omega)=\frac{I_n'(  \alpha\omega)}{K_{m}'(  \alpha\omega)} \frac{K_p'(  \beta\omega)}{I_p'(  \beta\omega)},\\
B_{mn;p}^{\text{DN}}(\omega)=&\frac{I_n(  \alpha\omega)}{K_{m}(  \alpha\omega)} \frac{K_p'(  \beta\omega)}{I_p'(  \beta\omega)},\hspace{1cm}
B_{mn;p}^{\text{ND}}(\omega)=\frac{I_n'(  \alpha\omega)}{K_{m}'(  \alpha\omega)} \frac{K_p(  \beta\omega)}{I_p(  \beta\omega)},\\
T_{mn;p}(\omega)=&I_{p-m}( (1-\vep) \omega)I_{p-n}((1-\vep)\omega).
\end{split}\end{equation}To find the first two leading terms of the Casimir interaction energy, one can replace the summations by integrations, i.e.,
\begin{equation}\label{eq7_29_6}
E_{\text{Cas}} \sim- \frac{L}{4\pi (b-a)^2}\sum_{s=0}^{\infty}\frac{1}{s+1}\int_0^{\infty}\omega \underbrace{\int_{ -\infty}^{\infty}\ldots\int_{ -\infty}^{\infty} }_{(2s+2) \;\text{times}}\prod_{i=0}^sA_{j_ij_{i+1};p_i}(\omega)\prod_{i=0}^s dp_i\prod_{i=0}^s dj_id\omega,
\end{equation}with the understanding that $j_{s+1}=j_0$.

Using Debye asymptotic expansions  of modified Bessel functions \cite{16}, one finds that $A_{\nu_1\nu_2;\nu_3}(\omega)$ has an expansion of the form
\begin{equation}\label{eq7_29_8}\begin{split}
A_{\nu_1\nu_2;\nu_3}(\omega)=&\pm\frac{\mathcal{A}}{2\pi}\exp\Bigl(\nu_1\eta(\omega_1) +\nu_2\eta(\omega_2)-2\nu_3\eta(\omega_3)+\nu_4\eta(\omega_4)+\nu_5\eta(\omega_5)\Bigr)\\&\times
\left(1+\frac{z_a(t(\omega_1))}{\nu_1}+\frac{z_a(t(\omega_2))}{\nu_2}-2\frac{z_b(t(\omega_3))}{\nu_3}
 +\frac{u_1(t(\omega_4))}{\nu_4}+\frac{u_1(t(\omega_5))}{\nu_5}\right),\end{split}\end{equation}where
\begin{align*}
&\nu_4=\nu_3-\nu_1,\quad\nu_5=\nu_3-\nu_2,\\
&\omega_1=\frac{\alpha\omega}{\nu_1},\quad\omega_2=\frac{\alpha\omega}{\nu_2},\quad\omega_3=\frac{\beta\omega}{\nu_3},\quad\omega_4=\frac{(1-\vep)\omega}{\nu_4},\quad
\nu_5=\frac{(1-\vep)\omega}{\nu_5};
\end{align*}
\begin{align*}
\eta(z)=\sqrt{1+z^2}+\ln\frac{z}{1+\sqrt{1+z^2}},\hspace{1cm}t(z)=\frac{1}{\sqrt{1+z^2}};
\end{align*}
\begin{equation*}\begin{split}
\mathcal{A}^{\text{DD}}=&\mathcal{A}^{\text{DN}}=\sqrt{\frac{\nu_1}{\nu_2\nu_4\nu_5}}
\left(\frac{1+\omega_1^2}{(1+\omega_2^2)(1+\omega_4^2)(1+\omega_5^2)}\right)^{\frac{1}{4}},\\
\mathcal{A}^{\text{NN}}=&\mathcal{A}^{\text{ND}}=\sqrt{\frac{\nu_2}{\nu_1\nu_4\nu_5}}
\left(\frac{1+\omega_2^2}{(1+\omega_1^2)(1+\omega_4^2)(1+\omega_5^2)}\right)^{\frac{1}{4}};
\end{split}\end{equation*}$z_c(t)=u_1(t)$ (or $z_c(t)=v_1(t)$) if the cylinder of radius $c$ is imposed with Dirichlet boundary condition (or Neumann boundary condition), with
\begin{align*}
u_1(t)=-\frac{5t^3-3t}{24},\hspace{1cm}v_1(t)=\frac{7t^3-9t}{24}.
\end{align*}The plus or minus sign in \eqref{eq7_29_8} depends on the boundary conditions. For DD or NN boundary conditions, we have the plus sign; whereas for DN or ND boundary conditions, we have the minus sign.

When the variables $\nu_1,\nu_2,\nu_3,\omega,\vep$ vary, the function
$$\nu_1\eta(\omega_1) +\nu_2\eta(\omega_2)-2\nu_3\eta(\omega_3)+\nu_4\eta(\omega_4)+\nu_5\eta(\omega_5)$$ is always nonpositive. It achieves the maximum value of $0$ when
$\nu_1=\nu_2, \vep=0$ and $$\nu_3=\frac{\beta}{\alpha}\nu_1=\frac{\beta}{\alpha}\nu_2.$$In that case,
$$\omega_1=\omega_2=\omega_3=\omega_4=\omega_5=\frac{\alpha\omega}{\nu_1}.$$
This suggests that in \eqref{eq7_29_5}, we can rename $j_0$ as $m$, and introduce new variables $n_1,\ldots, n_s$, $q_0,\ldots,q_s$ so that
\begin{equation*}\begin{split}&j_i=m+n_i,\quad 1\leq i\leq s,\\& p_i=\frac{\beta}{2\alpha}(2m+n_i+n_{i+1})+q_i,\quad 0\leq i\leq s.\end{split}\end{equation*} By a further substitution
$$\omega=\frac{m\sqrt{1-\tau^2}}{\alpha\tau},$$ we find that
\begin{equation}\label{eq7_29_10}
\begin{split}
E_{\text{Cas}}
\sim&-\frac{L}{2\pi a^2}\sum_{s=0}^{\infty}\frac{1}{s+1}\int_0^{1}  \int_{ 0}^{\infty}m^2\underbrace{\int_{ -\infty}^{\infty}\ldots\int_{ -\infty}^{\infty}}_{(2s+1)\;\text{times}}  \prod_{i=0}^{s}A_{m+n_i,m+n_{i+1};p_i}\left(\omega(\tau)\right)\prod_{i=0}^s dq_i\prod_{i=1}^s dn_i dm\frac{d\tau}{\tau^3},
\end{split}
\end{equation}with the understanding that $n_0=n_{s+1}=0$.

With
\begin{equation*}
\nu_1=m+n_i,\quad n_2=m+n_{i+1},\quad \nu_3=\frac{\beta}{2\alpha}(2m+n_i+n_{i+1})+q_i,
\end{equation*}
and treating  $n_i,n_{i+1},q_i$ and $\vep$ as perturbed variables, one can easily deduce that
$\nu_1\eta(\omega_1) +\nu_2\eta(\omega_2)-2\nu_3\eta(\omega_3)+\nu_4\eta(\omega_4)+\nu_5\eta(\omega_5)$ has an perturbative expansion of the form
\begin{equation}\label{eq7_29_7}
\nu_1\eta(\omega_1) +\nu_2\eta(\omega_2)-2\nu_3\eta(\omega_3)+\nu_4\eta(\omega_4)+\nu_5\eta(\omega_5)\sim\sum_{k=0}^{\infty}\sum_{j=0}^{\infty}\vep^jm^{1-k}\mathcal{G}_{kj}(n_i,n_{i+1},q_i)\mathcal{H}_{kj}(\tau),
\end{equation}
where $\mathcal{G}_{kj}(n_i,n_{i+1},q_i)$ is a homogeneous polynomial of degree $k$ in $n_i,n_{i+1},q_i$. The terms with $(k,j)=(0,0)$ and $(k,j)=(1,0)$ are identically zero. The leading terms are $-\mathfrak{M}_i$, where
\begin{equation*} \mathfrak{M}_i= \underbrace{\frac{2\vep m}{\alpha\tau}}_{(k,j)=(0,1)} +\underbrace{\frac{\beta\tau}{4m  }\left(n_i-n_{i+1}\right)^2
+\frac{\alpha^2 q_i^2\tau}{m\beta }}_{(k,j)=(2,0)}.
\end{equation*}comes from the terms with $(k,j)=(0,1)$ and $(k,j)=(2,0)$. This implies that the leading contribution to the Casimir interaction energy comes from $m\sim \vep^{-1}$, $n_i, n_{i+1}, q_i\sim \sqrt{m}\sim \vep^{-\frac{1}{2}}$.
Hence, the $(k,j)$ term in \eqref{eq7_29_7} is of order $\vep^{j+k/2-1}$. To keep everything up to order $\vep$, we need the terms in \eqref{eq7_29_7} with $j+k/2\leq 2$. With the help of a computer, we find that up to terms of order $\vep$,
\begin{equation*}
\nu_1\eta(\omega_1) +\nu_2\eta(\omega_2)-2\nu_3\eta(\omega_3)+\nu_4\eta(\omega_4)+\nu_5\eta(\omega_5)\sim -\mathfrak{M}_i+\mathfrak{A}_i+\mathfrak{B}_i,
\end{equation*}where $\mathfrak{A}_i$ and $\mathfrak{B}_i$ are terms of order $\sqrt{\vep}$ and $\vep$ given respectively by
\begin{equation*}\begin{split}
\mathfrak{A}_i=&\frac{\tau^3}{m^2}\left(\frac{\alpha^3(\alpha+2) q_i^3}{3 \beta^2}+\frac{\alpha^2 q_i^2(n_i+n_{i+1})}{2 \beta}
+\frac{\alpha^2q_i }{4  }\left(n_i-n_{i+1}\right)^2+\frac{\beta (n_i+n_{i+1})(n_i-n_{i+1})^2}{8  }\right)-\vep\tau\left(2 q_i+ \frac{ (n_i+n_{i+1})}{\alpha}\right),\\
\mathfrak{B}_i=&-\frac{\tau^3(3\tau^2-1)}{m^3}\Biggl(\frac{\alpha^4(\alpha^2+3\alpha+3) q_i^4}{12  \beta^3}+\frac{\alpha^3(\alpha+2) (n_i+n_{i+1})q_i^3}{6 \beta^2}
 +\frac{\alpha^2 q_i^2\left((\alpha^2+\alpha)(n_i-n_{i+1})^2+(n_i+n_{i+1})^2\right)}{8 \beta}
\\&+\frac{\alpha^2 q_i(n_i+n_{i+1}) (n_i-n_{i+1})^2}{8 }+\frac{ \beta}{192 }(n_i-n_{i+1})^2\Bigl((\alpha^2-\alpha)(n_i-n_{i+1})^2+7n_i^2+10n_in_{i+1}+7n_{i+1}^2\Bigr)\Biggr)\\
&-\frac{\vep  \tau(1-\tau^2)}{m}\left( \alpha q_i^2+    q_i  (n_i+n_{i+1})+\frac{\left(\alpha^2(n_i-n_{i+1})^2+(n_i+n_{i+1})^2\right)}{4\alpha  }\right)-\frac{\vep^2m\tau}{\alpha}.
\end{split}\end{equation*}In the same way, for $\mathcal{A}$, one finds that
\begin{equation*}
\mathcal{A}\sim   \frac{\alpha\tau}{m}\left(1+\mathfrak{C}_i+\mathfrak{D}_i\right),
\end{equation*}where $\mathfrak{C}_i$ and $\mathfrak{D}_i$ are respectively terms of order $\sqrt{\vep}$ and $\vep$ given  by
\begin{equation}\label{eq7_29_9}\begin{split}
\mathfrak{C}_i^{\text{DD}}=&-\frac{ \tau^2}{m}\left(\alpha q_i+n_{i+1}\right),\\
\mathfrak{D}_i^{\text{DD}}=&\vep(1-\tau^2)+\frac{ \alpha^2q_i^2\tau^2(3\tau^2-1)}{2m^2} -\frac{\alpha q_i \tau^2}{2m^2}\Bigl( (n_i+n_{i+1})-2\tau^2(  n_i+2 n_{i+1} )\Bigr)\\&
-\frac{\tau^2}{8m^2}\Bigl( \alpha^2(1-2\tau^2) (n_i-n_{i+1})^2+2\tau^2(n_i^2-2n_in_{i+1}-5n_{i+1}^2) -n_i^2+2n_in_{i+1}+3n_{i+1}^2\Bigl),
\end{split}\end{equation}in the DD case.
The DN case is the same as the DD case, and the NN or ND case can be obtained from \eqref{eq7_29_9} by interchanging $n_i$ and $n_{i+1}$.
Finally, one can check that the leading term of $$\frac{z_a(t(\omega_1))}{\nu_1}+\frac{z_a(t(\omega_2))}{\nu_2}-2\frac{z_b(t(\omega_3))}{\nu_3}
 +\frac{u_1(t(\omega_4))}{\nu_4}+\frac{u_1(t(\omega_5))}{\nu_5}$$ is of order $\vep$. Denote by $\mathfrak{F}_i$ the leading term. In the DD case, it is given by
\begin{equation*}
\mathfrak{F}_i^{\text{DD}}=-\frac{(1+\alpha+\alpha^2)\tau(5\tau^2-3)}{12m\beta},
\end{equation*}whereas for the NN, DN and ND case, we have
\begin{equation*}\begin{split}
\mathfrak{F}_i^{\text{NN}}=&\mathfrak{F}_i^{\text{DD}}+\frac{ \tau( \tau^2-1)}{m\beta},\\
\mathfrak{F}_i^{\text{DN}}=&\mathfrak{F}_i^{\text{DD}}-\frac{ \alpha\tau( \tau^2-1)}{\beta m},\\
\mathfrak{F}_i^{\text{ND}}=&\mathfrak{F}_i^{\text{DD}}+\frac{  \tau( \tau^2-1)}{  m}.
\end{split}\end{equation*}Notice that $\mathfrak{F}_i$ does not depend on $n_i, n_{i+1}$ and $q_i$.
Collecting the terms, we find that
\begin{equation*}\begin{split}
A_{m+n_i,m+n_{i+1};p_i}(\omega(\tau))\sim &\pm \frac{\alpha\tau}{2\pi m} \exp\Bigl(-\mathfrak{M}_i+\mathfrak{A}_i+\mathfrak{B}_i\Bigr)\Bigl(1+\mathfrak{C}_i+\mathfrak{D}_i\Bigr)\Bigl(1+\mathfrak{F}_i\Bigr)\\
\sim & \pm \frac{\alpha\tau}{2\pi m} e^{-\mathfrak{M}_i}\Bigl(1+\mathfrak{G}_i+\mathfrak{H}_i\Bigr),
\end{split}
\end{equation*}where $\mathfrak{G}_i$ and $\mathfrak{H}_i$ are respectively terms of order $\sqrt{\vep}$ and $\vep$ given by
\begin{equation*}
\begin{split}
\mathfrak{G}_i=&\mathfrak{A}_i+\mathfrak{C}_i,\\
\mathfrak{H}_i=&\frac{1}{2}\mathfrak{A}_i^2+\mathfrak{A}_i\mathfrak{C}_i+\mathfrak{B}_i+\mathfrak{D}_i+\mathfrak{F}_i.
\end{split}
\end{equation*}Substituting into the Casimir interaction energy \eqref{eq7_29_10}, we find that  up to first order correction term,
\begin{equation*}
\begin{split}
E_{\text{Cas}}
\sim&-\frac{\alpha^{s+1}L}{2^{s+2}\pi^{s+2} a^2}\sum_{s=0}^{\infty}\frac{(-1)^{\chi (s+1)}}{s+1}\int_0^{1}  \tau^{s-2}\int_{ 0}^{\infty}m^{1-s}\underbrace{\int_{ -\infty}^{\infty}\ldots\int_{ -\infty}^{\infty} }_{(2s+1) \;\text{times}}\exp\left(-\sum_{i=0}^s\mathfrak{M}_i\right)\\
&\hspace{4cm}\times\left(1+\sum_{i=0}^s\mathfrak{G}_i+\sum_{i=0}^{s-1}\sum_{j=i+1}^s\mathfrak{G}_i\mathfrak{G}_j+\sum_{i=0}^s\mathfrak{H}_i\right)\prod_{i=0}^sdq_i
 \prod_{i=1}^sdn_i dm d\tau.
\end{split}
\end{equation*}Here $\chi=0$ for DD or NN boundary conditions, and $\chi=1$ for DN or ND boundary conditions. The leading order term of the Casimir interaction energy  is
\begin{equation}\label{eq7_29_11}\begin{split}
E_{\text{Cas}}^0=
 &-\frac{\alpha^{s+1}L}{2^{s+2}\pi^{s+2} a^2}\sum_{s=0}^{\infty}\frac{(-1)^{\chi (s+1)}}{s+1}\int_0^{1}  \tau^{s-2}\int_{ 0}^{\infty}m^{1-s}\underbrace{\int_{ -\infty}^{\infty}\ldots\int_{ -\infty}^{\infty} }_{(2s+1) \;\text{times}}\exp\left(-\sum_{i=0}^s\mathfrak{M}_i\right)\prod_{i=0}^sdq_i
 \prod_{i=1}^sdn_i  dm d\tau.
\end{split}\end{equation}
The integrations of the term
$$\sum_{i=0}^s\mathfrak{G}_i$$ of order $\sqrt{\vep}$ over $n_i,n_{i+1},q_i$   give  zero since this term is odd in either $q_i, n_i$ or $n_{i+1}$. Thus the next-to-leading order term is
\begin{equation}\label{eq7_29_12}
\begin{split}
E_{\text{Cas}}^1=&-\frac{\alpha^{s+1}L}{2^{s+2}\pi^{s+2} a^2}\sum_{s=0}^{\infty}\frac{(-1)^{\chi (s+1)}}{s+1}\int_0^{1}  \tau^{s-2}\int_{ 0}^{\infty}m^{1-s}\underbrace{\int_{ -\infty}^{\infty}\ldots\int_{ -\infty}^{\infty} }_{(2s+1) \;\text{times}}\exp\left(-\sum_{i=0}^s\mathfrak{M}_i\right)\\
&\hspace{6cm}\times\left(  \sum_{i=0}^{s-1}\sum_{j=i+1}^s\mathfrak{G}_i\mathfrak{G}_j+\sum_{i=0}^s\mathfrak{H}_i\right)\prod_{i=0}^sdq_i
 \prod_{i=1}^sdn_i dm d\tau,
\end{split}
\end{equation}which is of order $\vep$ smaller than the leading order term.

Let us first consider the leading order term \eqref{eq7_29_11}. It  is straightforward to compute. Integrating over $q_i, 0\leq i\leq s$, first, we have
\begin{equation*}
\underbrace{\int_{-\infty}^{\infty}\ldots\int_{-\infty}^{\infty}}_{(s+1)\;\text{times}}\exp\left(-\sum_{i=0}^s\frac{\alpha^2 q_i^2\tau}{m\beta }\right)\prod_{i=0}^s dq_i
=\frac{\pi^{\frac{s+1}{2}}m^{\frac{s+1}{2}}\beta^{\frac{s+1}{2}}}{\alpha^{s+1}\tau^{\frac{s+1}{2}}}.
\end{equation*}For the integrations over $n_i,1\leq i\leq s$, since \cite{10}:
\begin{equation}\label{eq8_1_1}
\sum_{i=0}^s (n_i-n_{i+1})^2=2\left(n_1-\frac{1}{2}n_2\right)^2+\frac{3}{2}\left(n_2-\frac{2}{3}n_3\right)^2+\ldots+\frac{s+1}{s}n_s^2,
\end{equation} and
\begin{equation}\label{eq8_1_3}\int_{-\infty}^{\infty}e^{-\lambda (x-x_0)^2}dx=\sqrt{\frac{\pi}{\lambda}}\quad\text{for any $x_0$},\end{equation} integrating in the order $n_1\rightarrow n_2\rightarrow \ldots\rightarrow n_s$ gives
\begin{equation*}
\underbrace{\int_{-\infty}^{\infty}\ldots\int_{-\infty}^{\infty}}_{s\;\text{times}}\exp\left(-\sum_{i=0}^s\frac{\beta\tau}{4m  }\left(n_i-n_{i+1}\right)^2\right)\prod_{i=1}^s dn_i
=\frac{2^s\pi^{\frac{s}{2}}m^{\frac{s}{2}} }{\beta^{\frac{s}{2}}\tau^{\frac{s}{2}}\sqrt{s+1}}.
\end{equation*}Therefore,
\begin{equation}\label{eq7_29_13}\begin{split}
E_{\text{Cas}}^0=
 &-\frac{ \sqrt{\beta}L}{4\pi^{\frac{3}{2}} a^2}\sum_{s=0}^{\infty}\frac{(-1)^{\chi (s+1)}}{(s+1)^{\frac{3}{2}}}\int_0^{1}  \tau^{-\frac{5}{2}}\int_{ 0}^{\infty}m^{\frac{3}{2}} \exp\left(-\frac{2(s+1)\vep m}{\alpha\tau}\right)  dm d\tau\\
 =&-\frac{3\alpha^{\frac{5}{2}} \sqrt{\beta}L}{64\sqrt{2}\pi a^2\vep^{\frac{5}{2}}}\sum_{s=0}^{\infty}\frac{(-1)^{\chi (s+1)}}{(s+1)^{4}}\int_0^{1}  d\tau.
\end{split}\end{equation}Using the fact that
\begin{equation*}
\begin{split}
\sum_{s=0}^{\infty}\frac{1}{(s+1)^4}=\frac{\pi^4}{90},\hspace{1cm}\sum_{s=0}^{\infty}\frac{(-1)^s}{(s+1)^4}=\frac{7}{8}\frac{\pi^4}{90}=\frac{7\pi^4}{720},
\end{split}
\end{equation*}we find that for DD or NN boundary conditions, the leading order term of the Casimir interaction energy is
\begin{equation}\label{eq7_29_14}\begin{split}
E_{\text{Cas}}^0=&-\frac{\pi^3\sqrt{ab}L}{1920\sqrt{2(b-a)}d^{\frac{5}{2}}};
\end{split}\end{equation}whereas for DN or ND boundary conditions,
\begin{equation}\label{eq7_29_15}\begin{split}
E_{\text{Cas}}^0=&\frac{7\pi^3\sqrt{ab}L}{15360\sqrt{2(b-a)}d^{\frac{5}{2}}}.
\end{split}\end{equation}For the Casimir interaction force, one then obtains
\begin{equation}\label{eq7_29_16}
F_{\text{Cas}}^0=-\frac{ \pi^3\sqrt{ab}L}{768\sqrt{2(b-a)}d^{\frac{7}{2}}}
\end{equation}for DD or NN boundary conditions, and
\begin{equation}\label{eq7_29_17}
F_{\text{Cas}}^0=\frac{7\pi^3\sqrt{ab}L}{6144\sqrt{2(b-a)}d^{\frac{7}{2}}}
\end{equation}for DN or ND boundary conditions. The leading terms of the Casimir interaction forces \eqref{eq7_29_16} and \eqref{eq7_29_17} agree completely with that derive from proximity force approximations.

For the next-to-leading order term \eqref{eq7_29_12}, we integrate first with respect to $q_i$ using the formulas
\begin{equation}\label{eq8_1_4}
\int_{-\infty}^{\infty}q^{j}e^{-\lambda q^2}dq=\left\{\begin{aligned}&0,\hspace{2cm}&\text{if}\quad \text{$j$ is odd},\\&
\frac{\Gamma\left(\frac{j+1}{2}\right)}{\lambda^{\frac{j+1}{2}}},&\text{if}\quad \text{$j$ is even}.\end{aligned}\right.
\end{equation}
A straightforward computation gives
\begin{equation}\label{eq7_29_18}
\begin{split}
E_{\text{Cas}}^1=&-\frac{\beta^{\frac{s+1}{2}}L}{2^{s+2}\pi^{\frac{s+3}{2}} a^2}\sum_{s=0}^{\infty}\frac{(-1)^{\chi (s+1)}}{s+1}\int_0^{1}  \tau^{\frac{s-5}{2}}\int_{ 0}^{\infty}m^{\frac{3-s}{2}}\underbrace{\int_{ -\infty}^{\infty}\ldots\int_{ -\infty}^{\infty} }_{s \;\text{times}}\exp\left(-\frac{2(s+1)\vep m}{\alpha\tau}-\sum_{i=0}^s\frac{\beta\tau}{4m  }\left(n_i-n_{i+1}\right)^2\right)\\
&\hspace{6cm}\times\left(  \sum_{i=0}^{s-1}\sum_{j=i+1}^s\hat{\mathfrak{G}}_i\hat{\mathfrak{G}}_j+\sum_{i=0}^s\hat{\mathfrak{H}}_i\right)
 \prod_{i=1}^sdn_i dm d\tau,
\end{split}
\end{equation}where
\begin{equation*}\begin{split}
\hat{\mathfrak{G}}_i=&\frac{\alpha\sqrt{\tau}}{\sqrt{\pi m\beta}}\int_{-\infty}^{\infty}\exp\left(-\frac{\alpha^2 q_i^2\tau}{m\beta }\right)\mathfrak{G}_idq_i,\\
\hat{\mathfrak{H}}_i=&\frac{\alpha\sqrt{\tau}}{\sqrt{\pi m\beta}}\int_{-\infty}^{\infty}\exp\left(-\frac{\alpha^2 q_i^2\tau}{m\beta }\right)\mathfrak{H}_idq_i.
\end{split}\end{equation*}
The term $\hat{\mathfrak{H}}_i$ can be decomposed into two parts:
$$\hat{\mathfrak{H}}_i=\mathfrak{K}_i+\mathfrak{F}_i,$$where $\mathfrak{F}_i$ is as before and is independent of $n_i$ and $n_{i+1}$.
The explicit expressions for $\hat{\mathfrak{G}}_i$ and $ \mathfrak{K}_i$ are given in Appendix \ref{a1}. They are functions of $n_i$ and $n_{i+1}$. Let
\begin{equation*}
\begin{split}
I_{i}^s=&\frac{\beta^{\frac{s}{2}}\tau^{\frac{s}{2}}\sqrt{s+1}}{2^s\pi^{\frac{s}{2}}m^{\frac{s}{2}}}\underbrace{\int_{ -\infty}^{\infty}\ldots\int_{ -\infty}^{\infty} }_{s \;\text{times}}\exp\left( -\sum_{i=0}^s\frac{\beta\tau}{4m  }\left(n_i-n_{i+1}\right)^2\right)  \hat{\mathfrak{H}}_i
 \prod_{i=1}^sdn_i, \\
J_{ij}^s=& \frac{\beta^{\frac{s}{2}}\tau^{\frac{s}{2}}\sqrt{s+1}}{2^s\pi^{\frac{s}{2}}m^{\frac{s}{2}}}\underbrace{\int_{ -\infty}^{\infty}\ldots\int_{ -\infty}^{\infty} }_{s \;\text{times}}\exp\left( -\sum_{i=0}^s\frac{\beta\tau}{4m  }\left(n_i-n_{i+1}\right)^2\right)  \hat{\mathfrak{G}}_i\hat{\mathfrak{G}}_j
 \prod_{i=1}^sdn_i,
\end{split}
\end{equation*}and $$\mathcal{B}^s=\sum_{i=0}^s I_i^s+\sum_{i=0}^{s-1}\sum_{j=i+1}^s J_{ij}^s,$$so that
\begin{equation}\label{eq8_1_7}
\begin{split}
E_{\text{Cas}}^1=&-\frac{\sqrt{\beta}L}{4\pi^{\frac{3}{2}} a^2}\sum_{s=0}^{\infty}\frac{(-1)^{\chi (s+1)}}{(s+1)^{\frac{3}{2}}}\int_0^{1}  \tau^{-\frac{5}{2}}\int_{ 0}^{\infty}m^{\frac{3 }{2}} \exp\left(-\frac{2(s+1)\vep m}{\alpha\tau} \right)\mathcal{B}^s dm d\tau.
\end{split}
\end{equation}

The integrals $I_{i}^s$ and $J_{ij}^s$ are Gaussian and can   be integrated in exactly the same way as   in \cite{10}, which we explain in Appendix \ref{b1}. For the DD case,
we find that\begin{equation*}
\begin{split}
\mathcal{B}^{s,\text{DD}}=&\frac{\vep^2m\tau (s+1)}{3\alpha^2\beta }\left( (s+1)^2+3\alpha+2\right)\\&+\frac{\vep }{6\alpha\beta}\left(\left[(s+1)^2+(3\alpha+2)\right]\tau^2+\left[-2(s+1)^2+3\alpha^2-1\right]\right)\\
& + \frac{\tau\left([-7(s+1)^2+3\alpha+2]\tau^2+ 4(s+1)^2+\alpha^2   - \alpha   -1\right)}{16\beta m(s+1)}.
\end{split}
\end{equation*}For the DN case, since $\hat{\mathfrak{G}}_i^{\text{DN}}=\hat{\mathfrak{G}}_i^{\text{DD}}$ and
$$\hat{\mathfrak{H}}_i^{\text{DN}}=\hat{\mathfrak{H}}_i^{\text{DD}}+\mathfrak{F}_i^{\text{DN}}-\mathfrak{F}_i^{\text{DD}}=\hat{\mathfrak{H}}_i^{\text{DD}}-\frac{ \alpha\tau( \tau^2-1)}{\beta m},$$we find that
\begin{equation*}
\mathcal{B}^{s,\text{DN}}=\mathcal{B}^{s,\text{DD}}-\frac{ \alpha (s+1)\tau( \tau^2-1)}{\beta m}.
\end{equation*}For the NN case or the ND case,  $\hat{\mathfrak{G}}_i^{\text{NN}}=\hat{\mathfrak{G}}_i^{\text{ND}}$ and $ \mathfrak{K}_i^{\text{NN}}= \mathfrak{K}_i^{\text{ND}}$ can be obtained from $\hat{\mathfrak{G}}_i^{\text{DD}}$ and $\mathfrak{K}_i^{\text{DD}}$ respectively by interchanging $n_i$ and $n_{i+1}$. As explained in our previous work \cite{11}, these imply that $J_{ij}^{s,\text{NN}}=J_{ij}^{s,\text{ND}}=J_{s-j,s-i}^{s,\text{DD}}$ and
\begin{equation*}
\begin{split}
I_i^{s,\text{NN}}=&I_{s-i}^{s,\text{DD}}+\mathfrak{F}_i^{\text{NN}}-\mathfrak{F}_i^{\text{DD}}=I_{s-i}^{s,\text{DD}}+\frac{ \tau( \tau^2-1)}{m\beta},\\
I_i^{s,\text{ND}}=&I_{s-i}^{s,\text{DD}}+\mathfrak{F}_i^{\text{ND}}-\mathfrak{F}_i^{\text{DD}}=I_{s-i}^{s,\text{DD}}+\frac{  \tau( \tau^2-1)}{  m}.
\end{split}
\end{equation*}Hence,
\begin{align}
\mathcal{B}^{s,\text{NN}}=&\mathcal{B}^{s,\text{DD}}+\frac{ (s+1)\tau( \tau^2-1)}{m\beta},\label{eq8_1_8}\\
\mathcal{B}^{s,\text{ND}}=&\mathcal{B}^{s,\text{DD}}+\frac{ (s+1)\tau( \tau^2-1)}{m }\label{eq8_1_9}.
\end{align}We notice that the change in $\mathcal{B}^s$ due to the change of boundary conditions only comes from the term involving $u_1(t)$ and $v_1(t)$ in the Debye asymptotic expansions of the modified Bessel functions and their derivatives.

Now, it is easy to perform the integration over $m$ and $\tau$ in \eqref{eq8_1_7}. We find that in the DD case,
\begin{equation}\label{eq8_1_10}\begin{split}
E_{\text{Cas}}^{1,\text{DD}}=&-\frac{3\sqrt{ab }L}{64\sqrt{2}\pi (b-a)^{\frac{3}{2}} d^{\frac{3}{2}}}\sum_{s=0}^{\infty}\frac{1}{(s+1)^4}\int_0^{1}
\Biggl( \frac{(24\alpha+16)\tau^2+7\alpha^2-\alpha-3}{12\alpha\beta}\Biggr)d\tau\\
=&-\frac{3\sqrt{ab }L}{64\sqrt{2}\pi (b-a)^{\frac{3}{2}} d^{\frac{3}{2}}}\sum_{s=0}^{\infty}\frac{1}{(s+1)^4}\left(\frac{7}{12}+\frac{7}{36\alpha\beta} \right)\\
=&-\frac{\pi^3 \sqrt{ab }L}{1920\sqrt{2(b-a) }    d^{\frac{3}{2}}}\left(\frac{7}{12(b-a)}+\frac{7(b-a)}{36ab} \right).
\end{split}\end{equation}In the NN case, \eqref{eq8_1_8} implies that
\begin{equation}\label{eq8_1_11}
\begin{split}
E_{\text{Cas}}^{1,\text{NN}}=&E_{\text{Cas}}^{1,\text{DD}}-\frac{\sqrt{\beta}L}{4\pi^{\frac{3}{2}} a^2}\sum_{s=0}^{\infty}\frac{1}{(s+1)^{\frac{3}{2}}}\int_0^{1}  \tau^{-\frac{5}{2}}\int_{ 0}^{\infty}m^{\frac{3 }{2}} \exp\left(-\frac{2(s+1)\vep m}{\alpha\tau} \right)\frac{ (s+1)\tau( \tau^2-1)}{m\beta} dm d\tau\\
=&-\frac{3\sqrt{ab }L}{64\sqrt{2}\pi (b-a)^{\frac{3}{2}} d^{\frac{3}{2}}}\sum_{s=0}^{\infty}\frac{1}{(s+1)^4}\left(\frac{7}{12}+\frac{7}{36\alpha\beta} -\frac{8}{9\alpha\beta}(s+1)^2\right)\\
=&-\frac{\pi^3 \sqrt{ab }L}{1920\sqrt{2(b-a) }    d^{\frac{3}{2}}}\left(\frac{7}{12(b-a)}+\left[\frac{7 }{36 }-\frac{40}{3\pi^2}\right]\frac{b-a}{ab} \right).
\end{split}
\end{equation}Here we have used the fact that
$$\sum_{s=0}^{\infty}\frac{1}{(s+1)^2}=\frac{\pi^2}{6}.$$For DN and ND boundary conditions, the summation over $s$ is alternating in sign. Using
$$\sum_{s=0}^{\infty}\frac{(-1)^s}{(s+1)^2}=\frac{\pi^2}{12},$$
we have
\begin{equation}\label{eq8_1_12}
\begin{split}
E_{\text{Cas}}^{1,\text{DN}}=& \frac{3\sqrt{ab }L}{64\sqrt{2}\pi (b-a)^{\frac{3}{2}} d^{\frac{3}{2}}}\sum_{s=0}^{\infty}\frac{(-1)^s}{(s+1)^4}\left(\frac{7}{12}+\frac{7}{36\alpha\beta} +\frac{8}{9 \beta}(s+1)^2\right)\\
=& \frac{7\pi^3 \sqrt{ab }L}{15360\sqrt{2(b-a) }    d^{\frac{3}{2}}}\left(\frac{7}{12(b-a)}+ \frac{7 }{36 }\frac{b-a}{ab}+\frac{160}{21\pi^2} \frac{1}{b} \right).
\end{split}
\end{equation}Finally, for the ND case, it is easy to find that
 \begin{equation}\label{eq8_1_13}
\begin{split}
E_{\text{Cas}}^{1,\text{ND}}=&  \frac{7\pi^3 \sqrt{ab }L}{15360\sqrt{2(b-a) }    d^{\frac{3}{2}}}\left(\frac{7}{12(b-a)}+ \frac{7 }{36 }\frac{b-a}{ab}-\frac{160}{21\pi^2} \frac{1}{a} \right).
\end{split}
\end{equation}

Combining with the leading terms, we find that the asymptotic expansions of the Casimir interaction energies are given, up to the first  corrections, by
\begin{equation}\label{eq8_2_5}
\begin{split}
E_{\text{Cas}}^{\text{DD}}\sim &-\frac{\pi^3 \sqrt{ab }L}{1920\sqrt{2(b-a) }    d^{\frac{5}{2}}}\left(1+d\left[\frac{7}{12(b-a)}+\frac{7(b-a)}{36ab}\right]+\ldots \right),\\
E_{\text{Cas}}^{\text{NN}}\sim &-\frac{\pi^3 \sqrt{ab }L}{1920\sqrt{2(b-a) }    d^{\frac{5}{2}}}\left(1+d\left[\frac{7}{12(b-a)}+\left(\frac{7 }{36 }-\frac{40}{3\pi^2}\right)\frac{b-a}{ab}\right]+\ldots \right),\\
E_{\text{Cas}}^{\text{DN}}\sim &\frac{7\pi^3 \sqrt{ab }L}{15360\sqrt{2(b-a) }    d^{\frac{5}{2}}}\left(1+d\left[\frac{7}{12(b-a)}+\frac{7(b-a)}{36ab}+\frac{160}{21\pi^2} \frac{1}{b} \right]+\ldots \right),\\
E_{\text{Cas}}^{\text{ND}}\sim &\frac{7\pi^3 \sqrt{ab }L}{15360\sqrt{2(b-a) }    d^{\frac{5}{2}}}\left(1+d\left[\frac{7}{12(b-a)}+\frac{7(b-a)}{36ab}-\frac{160}{21\pi^2} \frac{1}{a} \right]+\ldots \right).
\end{split}
\end{equation}
For the Casimir interaction forces, we then obtain
\begin{equation}\label{eq8_2_6}
\begin{split}
F_{\text{Cas}}^{\text{DD}}\sim &-\frac{\pi^3 \sqrt{ab }L}{768\sqrt{2(b-a) }    d^{\frac{7}{2}}}\left(1+d\left[\frac{7}{20(b-a)}+\frac{7(b-a)}{60ab}\right]+\ldots \right),\\
F_{\text{Cas}}^{\text{NN}}\sim &-\frac{\pi^3 \sqrt{ab }L}{768\sqrt{2(b-a) }    d^{\frac{7}{2}}}\left(1+d\left[\frac{7}{20(b-a)}+\left(\frac{7 }{60 }-\frac{8}{ \pi^2}\right)\frac{b-a}{ab}\right]+\ldots \right),\\
F_{\text{Cas}}^{\text{DN}}\sim &\frac{7\pi^3 \sqrt{ab }L}{6144\sqrt{2(b-a) }    d^{\frac{7}{2}}}\left(1+d\left[\frac{7}{20(b-a)}+\frac{7(b-a)}{60ab}+\frac{32}{7\pi^2} \frac{1}{b} \right]+\ldots \right),\\
F_{\text{Cas}}^{\text{ND}}\sim &\frac{7\pi^3 \sqrt{ab }L}{6144\sqrt{2(b-a) }    d^{\frac{7}{2}}}\left(1+d\left[\frac{7}{20(b-a)}+\frac{7(b-a)}{60ab}-\frac{32}{7\pi^2} \frac{1}{a} \right]+\ldots \right).
\end{split}
\end{equation}
From these, it is easy to see that if the smaller cylinder is imposed with Dirichlet boundary conditions (the DD or DN case), then the proximity force approximation underestimates the strength of the Casimir interaction force. However, if the smaller cylinder is imposed with Neumann boundary conditions, then the proximity force approximation may overestimates or underestimates the strength of the Casimir interaction force depending on the relative sizes of the two cylinders.
Numerically,
\begin{equation*}\begin{split}
&\frac{7}{20(b-a)}+\left(\frac{7 }{60 }-\frac{8}{ \pi^2}\right)\frac{b-a}{ab}= -\frac{0.6939(b-0.4985a)(b-2.0059a)}{ab(b-a)},\\
&\frac{7}{20(b-a)}+\frac{7(b-a)}{60ab}-\frac{32}{7\pi^2} \frac{1}{a} = -\frac{0.3465(b+0.1815a)(b-1.8549a)}{ab(b-a)}
\end{split}\end{equation*}
 Therefore there is a critical ratio of $b/a$ over which the proximity force approximation overestimates the strength of the Casimir interaction force.

An interesting limiting case to study is the cylinder-plate configuration which can be achieved by taking the limit $b\rightarrow\infty$. In this case, we find from \eqref{eq8_2_6} that the first two leading terms of the Casimir interaction force are given by
\begin{equation}\label{eq8_2_2}
\begin{split}
F_{\text{Cas}}^{\text{CP,DD}}\sim &-\frac{\pi^3 \sqrt{a  }L}{768\sqrt{2  }    d^{\frac{7}{2}}}\left(1+ \frac{7 }{60 }\frac{d}{a}+\ldots \right),\\
F_{\text{Cas}}^{\text{CP,NN}}\sim &-\frac{\pi^3 \sqrt{a  }L}{768\sqrt{2 }    d^{\frac{7}{2}}}\left(1+\left[\frac{7 }{60 }-\frac{8}{ \pi^2}\right]\frac{d}{a} +\ldots \right),\\
F_{\text{Cas}}^{\text{CP,DN}}\sim &\frac{7\pi^3 \sqrt{a  }L}{6144\sqrt{2  }    d^{\frac{7}{2}}}\left(1+ \frac{7 }{60 }\frac{d}{a}+\ldots \right),\\
F_{\text{Cas}}^{\text{CP,ND}}\sim &\frac{7\pi^3 \sqrt{a  }L}{6144\sqrt{2}    d^{\frac{7}{2}}}\left(1+ \left[ \frac{7 }{60 }-\frac{32}{7\pi^2} \right]\frac{d}{a} +\ldots \right).
\end{split}
\end{equation}The results for the DD case and the NN case have been obtained in \cite{10}. However, to the best of our knowledge, the results for the DN and the ND case have not been obtained before.

\section{Two parallel cylinders exterior to each other}\label{s5}

\begin{figure}[h]\centering
\epsfxsize=0.5\linewidth \epsffile{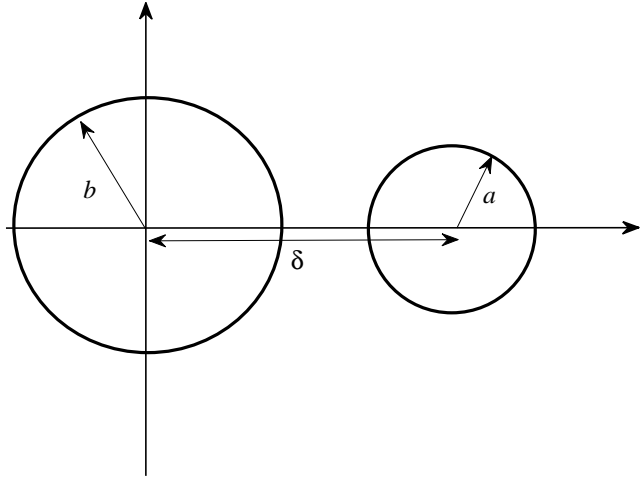}  \caption{\label{f2} The cross section of two  cylinders exterior to each other.}  \end{figure}

In this section, we   consider the case where two cylinders are parallel and exterior to each other (see Fig. \ref{f2}). The results for this case can be obtained in the same ways as   the case of one cylinder inside another cylinder which we consider above. Now, the distance between the centers of the cylinders $\delta$ is related to the radii $a$ and $b$ of the cylinders and the distance between the cylinders $d$ by $\delta=a+b-d$.
The proximity force approximation shows that at small separation (i.e., $d\ll a,b$), the leading term of the Casimir interacting force between the cylinders
is given by
\begin{equation} \begin{split}
F_{\text{Cas}}^{\text{PFA}}\sim & -\frac{ \pi^3 \sqrt{ab} L}{768\sqrt{2  (a+b)} d^{\frac{7}{2}}},
 \end{split}
\end{equation}for DD and NN boundary conditions. For DN and ND boundary conditions, we have
\begin{equation} \begin{split}
F_{\text{Cas}}^{\text{PFA}}\sim &  \frac{ 7\pi^3 \sqrt{ab} L}{6144\sqrt{2  (a+b)} d^{\frac{7}{2}}}.
 \end{split}
\end{equation}
The exact Casimir interaction energy between two cylinders which are exterior to each other has been derived using scattering approach or functional determinant method in \cite{17,4,5,18}. It can still be written in the form
\begin{equation}\label{eq8_2_1}
E_{\text{Cas}} = \frac{L}{4\pi}\int_0^{\infty}\xi \text{Tr}\ln \left(\mathbf{1}-\mathbf{M}(\xi)\right)d\xi,
\end{equation}where now the elements of the infinite matrix $\mathbf{M}$ are
\begin{equation}\label{eq8_2_3_1}\begin{split}
M_{mn}^{\text{DD}}(\xi)=&\frac{I_n(  a\xi)}{K_{m}(  a \xi)}\sum_{p=-\infty}^{\infty}\frac{I_p(  b \xi)}{K_p(  b\xi)}K_{p+m}( \delta \xi)K_{p+n}(\delta\xi)\\
M_{mn}^{\text{NN}}(\xi)=&\frac{I_n'(  a\xi)}{K_{m}'(  a \xi)}\sum_{p=-\infty}^{\infty}\frac{I_p'(  b \xi)}{K_p'(  b\xi)}K_{p+m}( \delta \xi)K_{p+n}(\delta\xi),\\
M_{mn}^{\text{DN}}(\xi)=&\frac{I_n(  a\xi)}{K_{m}(  a \xi)}\sum_{p=-\infty}^{\infty}\frac{I_p'(  b \xi)}{K_p'(  b\xi)}K_{p+m}( \delta \xi)K_{p+n}(\delta\xi),\\
M_{mn}^{\text{ND}}(\xi)=&\frac{I_n'(  a\xi)}{K_{m}'(  a \xi)}\sum_{p=-\infty}^{\infty}\frac{I_p(  b \xi)}{K_p(  b\xi)}K_{p+m}( \delta \xi)K_{p+n}(\delta\xi).
\end{split}\end{equation}
The computations  of the leading order term and the next-to-leading order correction term of the Casimir interaction energy go in parallel to the former case. Let $$\vep=\frac{d}{a+b}$$ be the dimensionless variable.  Then as $\vep\rightarrow 0^+$, we find that
\begin{equation}\label{eq8_2_4}
\begin{split}
E_{\text{Cas}}^{\text{DD}}\sim &-\frac{\pi^3 \sqrt{ab }L}{1920\sqrt{2(a+b) }    d^{\frac{5}{2}}}\left(1+d\left[-\frac{7}{12(a+b)}+\frac{7(a+b)}{36ab}\right]+\ldots \right),\\
E_{\text{Cas}}^{\text{NN}}\sim &-\frac{\pi^3 \sqrt{ab }L}{1920\sqrt{2(a+b) }    d^{\frac{5}{2}}}\left(1+d\left[-\frac{7}{12(a+b)}+\left(\frac{7 }{36 }-\frac{40}{3\pi^2}\right)\frac{a+b}{ab}\right]+\ldots \right),\\
E_{\text{Cas}}^{\text{DN}}\sim &\frac{7\pi^3 \sqrt{ab }L}{15360\sqrt{2(a+b) }    d^{\frac{5}{2}}}\left(1+d\left[-\frac{7}{12(a+b)}+\frac{7(a+b)}{36ab}-\frac{160}{21\pi^2} \frac{1}{b} \right]+\ldots \right),\\
E_{\text{Cas}}^{\text{ND}}\sim &\frac{7\pi^3 \sqrt{ab }L}{15360\sqrt{2(a+b) }    d^{\frac{5}{2}}}\left(1+d\left[-\frac{7}{12(a+b)}+\frac{7(a+b)}{36ab}-\frac{160}{21\pi^2} \frac{1}{a} \right]+\ldots \right).
\end{split}
\end{equation}
For the Casimir interaction force, we then obtain
\begin{equation}\label{eq8_2_3}
\begin{split}
F_{\text{Cas}}^{\text{DD}}\sim &-\frac{\pi^3 \sqrt{ab }L}{768\sqrt{2(a+b) }    d^{\frac{7}{2}}}\left(1+d\left[-\frac{7}{20(a+b)}+\frac{7(a+b)}{60ab}\right]+\ldots \right),\\
F_{\text{Cas}}^{\text{NN}}\sim &-\frac{\pi^3 \sqrt{ab }L}{768\sqrt{2(a+b) }    d^{\frac{7}{2}}}\left(1+d\left[-\frac{7}{20(a+b)}+\left(\frac{7 }{60 }-\frac{8}{ \pi^2}\right)\frac{a+b}{ab}\right]+\ldots \right),\\
F_{\text{Cas}}^{\text{DN}}\sim &\frac{7\pi^3 \sqrt{ab }L}{6144\sqrt{2(a+b) }    d^{\frac{7}{2}}}\left(1+d\left[-\frac{7}{20(a+b)}+\frac{7(a+b)}{60ab}-\frac{32}{7\pi^2} \frac{1}{b} \right]+\ldots \right),\\
F_{\text{Cas}}^{\text{ND}}\sim &\frac{7\pi^3 \sqrt{ab }L}{6144\sqrt{2(a+b) }    d^{\frac{7}{2}}}\left(1+d\left[-\frac{7}{20(a+b)}+\frac{7(a+b)}{60ab}-\frac{32}{7\pi^2} \frac{1}{a} \right]+\ldots \right).
\end{split}
\end{equation}
Notice that for the DD and the NN cases, there is a complete symmetry between the parameters $a$ and $b$. On the other hand, the ND case can be obtained from the DN case by interchanging the parameters $a$ and $b$. These are expected since in the present case, the two cylinders are on equal footing.

Numerically, $7/60-8/\pi^2=-0.6939$. Therefore, it is easy to conclude that in the NN case, the proximity force approximation overestimates the strength of the Casimir interaction  force. For the DD boundary conditions,
\begin{equation*}
\begin{split}
-\frac{7}{20(a+b)}+\frac{7(a+b)}{60ab}=\frac{7(a^2-ab+b^2)}{60ab(a+b)}>0.
\end{split}
\end{equation*}Therefore, the proximity force approximation underestimates the Casimir interaction force.
For the ND boundary conditions,
\begin{equation*}
\begin{split}
-\frac{7}{20(a+b)}+\frac{7(a+b)}{60ab}-\frac{32}{7\pi^2} \frac{1}{a} =-\frac{0.3465(b+1.8549a)(b-0.1815a)}{ab(a+b)}.
\end{split}
\end{equation*}Therefore, we find that the proximity force approximation may underestimate or overestimate the strength of the Casimir interaction force depending on the ratio of the two radii of the cylinders.

It is interesting to compare the results for two cylinders exterior to each other \eqref{eq8_2_4} and \eqref{eq8_2_3} with the results for one cylinder inside another cylinder \eqref{eq8_2_5} and \eqref{eq8_2_6}. Notice that they have similar coefficients up to the changes of signs. This can be considered as an analogy of the result obtained in \cite{51}, where the exact closed forms for the Casimir energy of two weakly coupled dielectric cylinders were derived and it was shown that the analytic results for one cylinder inside the other can be obtained as an analytic continuation of the result for two cylinders exterior to each other.

Taking the limit $b\rightarrow \infty$, we again recover the configuration of a cylinder in front of a plate. It is easy to check that taking the $b\rightarrow\infty$ limits of \eqref{eq8_2_3} give \eqref{eq8_2_2}.

\section{Conclusion}

In this article, we have computed analytically the asymptotic expansion of the Casimir interaction force between two cylinders, with one inside the another, or both  outside each other. We compute the leading order term and the next-to-leading order term. Different combinations of Dirichlet (D) and Neumann (N) boundary conditions are discussed. The results read as
\begin{equation}\label{eq8_4_1}
\begin{split}
F_{\text{Cas}}^{\text{DD}}\sim &-\frac{\pi^3 \sqrt{ab }L}{768\sqrt{2(b\mp a) }    d^{\frac{7}{2}}}\left(1+d\left[\pm\frac{7}{20(b\mp a)}+\frac{7(b\mp a)}{60ab}\right]+\ldots \right),\\
F_{\text{Cas}}^{\text{NN}}\sim &-\frac{\pi^3 \sqrt{ab }L}{768\sqrt{2(b\mp a) }    d^{\frac{7}{2}}}\left(1+d\left[\pm\frac{7}{20(b\mp a)}+\left(\frac{7 }{60 }-\frac{8}{ \pi^2}\right)\frac{b\mp a}{ab}\right]+\ldots \right),\\
F_{\text{Cas}}^{\text{DN}}\sim &\frac{7\pi^3 \sqrt{ab }L}{6144\sqrt{2(b\mp a) }    d^{\frac{7}{2}}}\left(1+d\left[\pm\frac{7}{20(b\mp a)}+\frac{7(b\mp a)}{60ab}\pm\frac{32}{7\pi^2} \frac{1}{b} \right]+\ldots \right),\\
F_{\text{Cas}}^{\text{ND}}\sim &\frac{7\pi^3 \sqrt{ab }L}{6144\sqrt{2(b\mp a) }    d^{\frac{7}{2}}}\left(1+d\left[\pm\frac{7}{20(b\mp a)}+\frac{7(b\mp a)}{60ab}-\frac{32}{7\pi^2} \frac{1}{a} \right]+\ldots \right).
\end{split}
\end{equation}For the terms $\pm$ or $\mp$, the sign on the top is for the case where one cylinder is inside another, and the sign at the bottom is for the case where the two cylinders are exterior to each other. It is observed that for each case,   the leading term of the Casimir interaction force agrees with that derived using the proximity force approximation. The proximity force approximation may underestimate or overestimate the magnitude of the Casimir interaction force depending on the boundary conditions and  the ratio of the radii. The special $b\rightarrow \infty$ limiting case  which gives the results for the cylinder-plate configuration  is discussed. It is found that the $b\rightarrow\infty$ limits of the asymptotic expansions for the DD and NN cases reproduce the well-known results for the corresponding asymptotic expansions for the cylinder-plane configuration.

Although we only consider the Dirichlet and Neumann boundary conditions in this article, it is easy to obtain from \eqref{eq8_4_1} the asymptotic expansions for perfectly conducting or infinitely permeable cylinders. More specifically, if both the   cylinders are perfectly conducting or both are infinitely permeable, one takes the sum of the results for the DD and the NN cases. If one cylinder is perfectly conducting and one is infinitely permeable, then one takes the sum of the results for the DN and the ND cases.

\appendix
\section{Explicit expressions for $\boldsymbol{\hat{\mathfrak{G}}_i}$ and $\boldsymbol{ \mathfrak{K}_i}$}\label{a1}
For $\hat{\mathfrak{G}}_i$,
\begin{equation*}
\begin{split}
\hat{\mathfrak{G}}_i^{\text{DD}}=\hat{\mathfrak{G}}_i^{\text{DN}}=\frac{  \tau^2(n_i-3n_{i+1})}{4m }
 +\frac{\beta\tau^3(n_i+n_{i+1})(n_i-n_{i+1})^2}{8m^2 }- \vep\frac{\tau(n_i+n_{i+1})}{\alpha},
\end{split}
\end{equation*}and $\hat{\mathfrak{G}}_i^{\text{NN}}$ and  $\hat{\mathfrak{G}}_i^{\text{ND}}$ are obtained from this by interchanging $n_i$ and $n_{i+1}$.
For $\mathfrak{K}_i$,
\begin{equation*}\begin{split}
 \mathfrak{K}_i^{\text{DD}}= \mathfrak{K}_i^{\text{DN}}=& -\frac{(\alpha^2+3\alpha+3)(3\tau^2-1)\tau }{16  m\beta} -\frac{\tau^2(3\tau^2-1)}{16m^2}\left((\alpha^2+\alpha)(n_i-n_{i+1})^2+(n_i+n_{i+1})^2\right)\\
&-\frac{\tau^3(3\tau^2-1)\beta}{192m^3}(n_i-n_{i+1})^2\Bigl((\alpha^2-\alpha)(n_i-n_{i+1})^2+7n_i^2+10n_in_{i+1}+7n_{i+1}^2\Bigr)\\
&-\frac{\vep \beta   (1-\tau^2)}{2 \alpha}-\frac{\vep \tau(1-\tau^2)}{4\alpha m}\left(\alpha^2(n_i-n_{i+1})^2+(n_i+n_{i+1})^2\right)+\vep(1-\tau^2)+\frac{ \beta \tau (3\tau^2-1)}{4m}  \\&
-\frac{\tau^2}{8m^2}\Bigl(-2\alpha^2\tau^2(n_i-n_{i+1})^2+2\tau^2(n_i^2-2n_in_{i+1}-5n_{i+1}^2)+\alpha^2(n_i-n_{i+1})^2-n_i^2+2n_in_{i+1}+3n_{i+1}^2\Bigl)\\
&+\frac{5(\alpha+2)^2\tau^3 }{48m\beta}+\frac{\alpha(\alpha+2)\tau^4 }{16m^2 }\left(n_i-n_{i+1}\right)^2-\frac{\vep (\alpha+2)\tau^2 }{2\alpha }
+\frac{  3\tau^4 (n_i+n_{i+1})^2}{32m^2 }+\frac{  \beta\tau^5 (n_i^2-n_{i+1}^2)^2}{32m^3 }\\
&-\frac{\vep  \tau^3 (n_i+n_{i+1})^2}{4m \alpha }+\frac{\alpha^2 \beta\tau^5}{64m^3 }\left(n_i-n_{i+1}\right)^4-\frac{\vep  \beta\tau^3}{4 m  }\left(n_i-n_{i+1}\right)^2
+\frac{\beta^2\tau^6(n_i+n_{i+1})^2(n_i-n_{i+1})^4}{128m^4 }\\&- \frac{\vep\beta\tau^4 (n_i^2-n_{i+1}^2)^2}{8m^2\alpha }
+ \vep^2\frac{ \tau m\beta}{\alpha^2}+\vep^2\frac{\tau^2(n_i+n_{i+1})^2}{2\alpha^2}-\frac{(\alpha+2)\tau^3 }{4m}-\frac{\alpha\beta \tau^4}{8m^2}\left(n_i-n_{i+1}\right)^2+\frac{\vep\beta\tau^2}{\alpha}\\&-\frac{ \tau^4n_{i+1}(n_i+n_{i+1})}{4m^2 } -\frac{\beta\tau^5n_{i+1}(n_i+n_{i+1})(n_i-n_{i+1})^2}{8m^3}+\vep\frac{\tau^3n_{i+1}(n_i+n_{i+1})}{m\alpha}-\frac{\vep^2m\tau}{\alpha}.
\end{split}\end{equation*}$\mathfrak{K}_i^{\text{NN}}$ and  $ \mathfrak{K}_i^{\text{ND}}$ are obtained from this by interchanging $n_i$ and $n_{i+1}$.

\section{The integrals $\boldsymbol{I_i^s}$ and $\boldsymbol{J_{ij}^s}$}\label{b1}
For $I_0^s$, using the identity \begin{equation}\label{eq8_1_5}
\sum_{i=0}^s(n_i-n_{i+1})^2=2\left(n_s-\frac{1}{2}n_{s-1}\right)^2+\frac{3}{2}\left(n_{s-1}-\frac{2}{3}n_{s-2}\right)^2+\ldots+\frac{s+1}{s}n_1^2,\end{equation}we can  integrate in the order $n_s\rightarrow\ldots\rightarrow n_2\rightarrow n_1$. The integrations over $n_s,\ldots,n_2$ only require the formula \eqref{eq8_1_3}, whereas the integration over $n_1$ needs the formulas  \eqref{eq8_1_4}.

For $I_s^s$, we use \eqref{eq8_1_1} and integrate in the order $n_1\rightarrow\ldots\rightarrow n_{s}$.

For $I_i^s,1\leq i\leq s-1$, using the identity
\begin{equation*}\begin{split}
\sum_{i=0}^s(n_i-n_{i+1})^2=&2\left(n_1-\frac{1}{2}n_2\right)^2+\ldots+\frac{i+1}{i}\left(n_{i }-\frac{i }{i+1}n_{i+1}\right)^2
+2\left(n_s-\frac{1}{2}n_{s-1}\right)^2+\ldots\\
 &+\frac{s-i}{s-i-1}\left(n_{i+2}-\frac{s-i-1}{s-i}n_{i+1}\right)^2+\frac{s +1}{(i+1)(s-i)}n_{i+1}^2,
\end{split}\end{equation*}we can first integrate in the order $n_1\rightarrow\ldots\rightarrow n_{i-1}\rightarrow n_s\rightarrow\ldots\rightarrow n_{i+2}$,  and then make a change of variables
$$x=n_{i }-\frac{i }{i+1}n_{i+1},\quad y=n_{i+1}.$$The integrations over $x$ and $y$ can be performed using \eqref{eq8_1_4}.

It is interesting to remark that although the cases of $I_{0}^s$ and $I_{s}^s$ have to be considered separately, it turns out that by formally substituting $i=0$ and $i=s$ into the formula obtained for $I_{i}^s$ where $1\leq i\leq s-1$, the results agree with $I_0^s$ and $I_s^s$ respectively.

For $J_{01}^s$, using the identity  \eqref{eq8_1_5}, we can  first integrate in the order $n_s\rightarrow\ldots\rightarrow n_3$. Then make a change of variables
$$x=n_1,\quad y=n_2-\frac{s-1}{s}n_1,$$and integrate over $x$ and $y$ using \eqref{eq8_1_4}.

For $J_{0s}^s$, using the identity
\begin{align*}
\sum_{i=0}^s(n_i-n_{i+1})^2
=&2\left(m_2-\frac{1}{2}m_{3}\right)^2+\ldots+ \frac{s-2}{s-3}\left(m_{s-2}-\frac{s-3}{s-2}m_{s-1}\right)^2+\frac{s-1}{s-2}\left(m_{s-1}-\frac{s-2}{s-1}(n_s-n_1)\right)^2\\
&+\frac{s}{s-1}\left(n_1-\frac{1}{s}n_s\right)^2+\frac{s+1}{s}n_s^2,
\end{align*}where $m_i=n_i-n_1,\;2\leq i\leq s-1$, we can first integrate in the order $m_2\rightarrow\ldots\rightarrow m_{s-1}$. Then make a change of variables
\begin{equation*}
x=n_1-\frac{1}{s}n_s,\quad y=n_s,
\end{equation*}and integrate over $x$ and $y$.

For $J_{0j}^s$, where $2\leq j\leq s-1$, using the identity
\begin{align*}
\sum_{i=0}^s(n_i-n_{i+1})^2= &2\left(m_2-\frac{1}{2}m_3\right)^2+\ldots+\frac{j-1}{j-2}\left(m_{j-1}-\frac{j-2}{j-1}(n_j-n_1)\right)^2\\&+2\left(n_s-\frac{1}{2}n_{s-1}\right)^2+\ldots+\frac{s-j}{s-j-1}
\left(n_{j+2}-\frac{s-j-1}{s-j}n_{j+1}\right)\\
&+\frac{j}{j-1}\left(n_1-\frac{1}{j}n_j\right)^2+\frac{s+1}{j(s-j+1)}n_j^2+\frac{s-j+1}{s-j}\left(n_{j+1}-\frac{s-j}{s-j+1}n_j\right)^2,
\end{align*}where $m_i=n_i-n_1,2\leq i\leq j-1$, we can first integrate in the order $m_2\rightarrow\ldots\rightarrow m_{j-1}\rightarrow n_s\rightarrow\ldots\rightarrow n_{j+2}$. Then make a change of variables
\begin{align*}
x=n_1-\frac{1}{j}n_j,\quad y=n_j,\quad z=n_{j+1}-\frac{s-j}{s-j+1}n_j,
\end{align*}and integrate over $x,y$ and $z$.

For $J_{s-1,s}^s$, using the identity
\eqref{eq8_1_1}, we can first integrate in the order $n_1\rightarrow\ldots\rightarrow  n_{s-2}$. Then make a change of variables
$$x=n_{s-1 }-\frac{s-1 }{s}n_s,\quad y=n_{s}$$ and integrate over $x$ and $y$.

For $J_{is}^s, 1\leq i\leq s-2$, using the identity
\begin{align*}
\sum_{i=0}^s(n_i-n_{i+1})^2=&2\left(n_1-\frac{1}{2}n_2\right)^2+\ldots+\frac{i}{i-1}\left(n_{i-1}-\frac{i-1}{i}n_i\right)^2\\
 &+2\left(m_{s-1}-\frac{1}{2}m_{s-2}\right)^2+\ldots+
\frac{s-i-1}{s-i-2}\left(m_{i+2}-\frac{s-i-2}{s-i-1}(n_{i+1}-n_s)\right)^2\\&
+\frac{i+1}{i}\left(n_i-\frac{i}{i+1}n_{i+1}\right)^2+\frac{s+1}{(i+1)(s-i)}n_{i+1}^2+\frac{s-i}{s-i-1}\left(n_s-\frac{1}{s-i}n_{i+1}\right)^2
\end{align*}where $  m_j=n_j-n_s,  i+2\leq j\leq s-1$, we can first integrate in the order $n_1\rightarrow\ldots\rightarrow n_{i-1}\rightarrow m_{s-1}\rightarrow\ldots\rightarrow m_{i+2}$. Then make a change of variables
$$x=n_i-\frac{i}{i+1}n_{i+1},\quad y=n_{i+1},\quad z=n_s-\frac{1}{s-i}n_{i+1}, $$ and integrate over $x,y$ and $z$.

For $J_{i,i+1}^s, 1\leq i\leq s-2$, using the identity
\begin{align*}
\sum_{i=0}^s(n_i-n_{i+1})^2
=&2\left(n_1-\frac{1}{2}n_2\right)^2+\ldots+\frac{i+1}{i}\left(n_{i}-\frac{i}{i+1}n_{i+1}\right)^2\\&
+2\left(n_s-\frac{1}{2}n_{s-1}\right)^2+\ldots+\frac{s-i}{s-i-1}\left(n_{i+2}-\frac{s-i-1}{s-i}n_{i+1}\right)^2
+\frac{s+1}{(i+1)(s-i)}n_{i+1}^2,
\end{align*}we can first integrate in the order $n_1\rightarrow\ldots\rightarrow n_{i-1}\rightarrow n_s\rightarrow\ldots\rightarrow n_{i+3}$. Then make a change of variables
\begin{align*}
&x=n_i-\frac{i}{i+1}n_{i+1},\quad y=n_{i+1},\quad z=n_{i+2}-\frac{s-i-1}{s-i}n_{i+1}
\end{align*}and integrate over $x,y$ and $z$.

Finally, for $J_{ij}^s$, where $1\leq i\leq s-3, i+2\leq j\leq s-1$, using the identity
\begin{align*}
\sum_{i=0}^s(n_i-n_{i+1})^2
=&2\left(n_1-\frac{1}{2}n_2\right)^2+\ldots+\frac{i+1}{i}\left(n_{i}-\frac{i}{i+1}n_{i+1}\right)^2\\&+2\left(n_s-\frac{1}{2}n_{s-1}\right)^2+\ldots+\frac{s-j+1}{s-j}\left(n_{j+1}-\frac{s-j}{s-j+1}n_{j}\right)^2
\\&+2\left(m_{i+2}-\frac{1}{2}m_{i+3}\right)^2+\ldots+\frac{j-i-1}{j-i-2}\left(m_{j-1}-\frac{j-i-2}{j-i-1}(n_j-n_{i+1})\right)^2\\
&+\frac{j}{(i+1)(j-i-1)}\left(n_{i+1}-\frac{i+1}{j}n_j \right)^2+\frac{s+1}{j(s-j+1)}n_j^2,
\end{align*}where $m_k=n_k-n_{i+1}, i+2\leq k\leq j-1$, we can  first integrate in the order $n_1\rightarrow\ldots\rightarrow n_{i-1}\rightarrow n_s\rightarrow\ldots\rightarrow n_{j+2}\rightarrow m_{i+2}\rightarrow \ldots\rightarrow m_{j-1}$. Then make a change of variables
\begin{align*}
&x=n_i-\frac{i}{i+1}n_{i+1}, \quad y=n_{i+1}-\frac{i+1}{j}n_j,\quad z=n_j,\quad w=n_{j+1}-\frac{s-j}{s-j+1}n_{j},
\end{align*}and integrate over $x,y,z$ and $w$.

As in the case of $I_i^s$, by formally substituting $(i,j)=(0,1), (0,s), (0,j), (s-1,s), (i,s), (i,i+1)$ into the expressions obtained for $J_{ij}^s$ where $1\leq i\leq s-3, i+2\leq j\leq s-1$, one obtains respectively $J_{01}^s, J_{0s}^s, J_{0j}^s, J_{s-1,s}^s, J_{is}^s, J_{i,i+1}^s$.
\begin{acknowledgments}
 This project is funded by the Ministry of Higher Education of Malaysia   under the FRGS grant FRGS/2/2010/SG/UNIM/02/2.
\end{acknowledgments}
 
\end{document}